\documentclass[twocolumn,aps,showpacs,showkeys,prb,superscriptaddress,reprint]{revtex4-1}
\usepackage{graphicx}
\usepackage{amsmath}
\usepackage{amssymb}
\usepackage{amsfonts}
\usepackage{color}
\usepackage{bm}
\usepackage{hyperref}

\hypersetup{
	colorlinks = true,
	pdftitle = {},
	pdfauthor = {},
	pdfkeywords = {},
	linkcolor = blue,
	citecolor = blue,
	filecolor = black,
	urlcolor = magenta
}

\newcommand{\Vacc}{V_\text{acc}}

\allowdisplaybreaks

\begin{document}

\title{Fully non-local inelastic scattering computations for spectroscopical TEM methods}

\author{J\'{a}n Rusz}
\email{jan.rusz@physics.uu.se}
\affiliation{Department of Physics and Astronomy, Uppsala University, P.O. Box 516, 75120 Uppsala, Sweden}
\author{Axel Lubk}
\affiliation{Leibniz Institute for Solid State and Materials Research, Helmholtzstra\ss{}e 20, 01069 Dresden, Germany}
\author{Jakob Spiegelberg}
\author{Dmitry Tyutyunnikov}
\affiliation{Department of Physics and Astronomy, Uppsala University, P.O. Box 516, 75120 Uppsala, Sweden}

\begin{abstract}
The complex interplay of elastic and inelastic scattering amenable to different levels of approximation constitutes the major challenge for the computation and hence interpretation of TEM-based spectroscopical methods. The two major approaches to calculate inelastic scattering cross-sections of fast electrons on crystals -- Yoshioka-equations-based forward propagation and reciprocal wave method -- are founded in two conceptually differing schemes -- a numerical forward integration of each inelastically scattered wave function, yielding the exit density matrix, and a computation of inelastic scattering matrix elements using elastically scattered initial and final states (double channeling). Here, we compare both approaches and show that the latter is computationally competitive to the former by exploiting analytical integration schemes over multiple excited states. Moreover, we show how to include full non-locality of the inelastic scattering event, neglected in the forward propagation approaches, at no additional computing costs in the reciprocal wave method. Detailed simulations show in some cases significant errors due to the $z$-locality approximation and hence pitfalls in the interpretation of spectroscopical TEM results.
\end{abstract}

\pacs{}
\keywords{}

\maketitle

\section{Introduction}

Simulation of elastic and inelastic scattering cross-sections of fast electrons on crystals is an indispensable tool for predicting and interpreting experimental findings in transmission electron microscopy (TEM). For inelastic electron scattering, the ultimate task is to evaluate the double-differential scattering cross-section (DDSCS) $\frac{\partial^2 \sigma}{\partial \Omega \partial E}$, which represents the likelihood of electrons to be scattered into a specific angle $\Omega$, while losing energy $E$. If necessary, such calculations should take into consideration the electron optical parameters, such as aberrations or the partial coherence of the electron source. This is particularly important for simulations at atomic resolution. In this work we will focus on simulations of core-level excitations, however, many of the conclusions remain valid for general inelastic processes, such as scattering on plasmons or phonons. Moreover, we will focus on coherent electron sources to simplify notation. The ramifications of a partially coherent illumination may be, however, incorporated straightforwardly by summing over the (incoherent) emitter size.

In the present state-of-the-art literature (as of early 2017) the most commonly used approach for inelastic DDSCS calculations is a multislice approach based on Yoshioka's equations \cite{yoshioka,wang,allen,dwyer,verbeeck,oxley,allen2}. In this approach the incoming electron wave, represented by the ket vector $|\psi_{z=0}\rangle$, is elastically propagated through the crystal until the inelastic scattering site (e.g., an atom). We introduce the notation $\hat{U}(z_2,z_1)$ for the $z$-evolution operator, which elastically propagates the wavefunction from its state at coordinate $z=z_1$ to $z=z_2$. This is typically implemented in the paraxial approximation (i.e., small scattering angles)

\begin{equation}
  \left| \frac{\partial^2 \varphi}{\partial z^2} \right| \ll \left| k_z \frac{\partial \varphi}{\partial z} \right|
\end{equation}
which, using an Ansatz $\psi(\mathbf{r}) = e^{ik_z z}\varphi(\mathbf{r})$, leads to a conventional multislice method \cite{cowley,kirkland}
\begin{equation}
  \frac{\partial}{\partial z}\varphi = i \left[ \frac{1}{2 k_z} \Delta_{\perp} + \sigma V \right]\varphi ,
\end{equation}
where $\Delta_{\perp}$ is a Laplacian in $x,y$ coordinates, $k_z$ is the wave-vector component along the beam direction, $\sigma$ is an interaction constant depending solely on the acceleration voltage,
and $V=V(\mathbf{r})$ is the crystal potential. This has exactly the form of a time-dependent 2-dimensional Schr\"{o}dinger equation, where the time is represented by the $z$-coordinate, which allows to explicitly construct an evolution operator in the $z$-coordinate
\begin{equation}
  \hat{U}(z_2,z_1) = \hat{Z} \exp \left\{ i \int_{z_1}^{z_2} \left[ \frac{1}{2 k_z} \Delta_{\perp} + \sigma V \right] dz \right\}, 
\end{equation}
This operator is unitary, i.e., $\hat{U}(z_2,z_1)\hat{U}(z_1,z_2)=\hat{1}$ and $\hat{U}(z_2,z_1) = \hat{U}^\dagger(z_1,z_2)$, because both $\Delta_{\perp}$ and $V(\mathbf{r})$ are hermitean. $\hat{Z}$ is Dyson's $z$-ordering operator.

To take into account the incoherency between different (orthogonal) inelastic events, a new computational thread must be spawned, whenever the slice contains an atom, for which the core-level excitations fall into the energy range of the DDSCS, which we want to evaluate. A transition potential (M\o{}ller potential) is evaluated for an atom located at $\mathbf{a}$, describing its excitation from a many-electron ground state $|i_\mathbf{a}\rangle$ to a specific many-electron final state $|f_\mathbf{a}\rangle$, which modifies the beam wavefunction from $|\psi_{z=z_a}^{i_\mathbf{a}}\rangle = \hat{U}(z_a,0)|\psi_{z=0}\rangle$ to $|\phi_{z=z_a}^{f_\mathbf{a}}\rangle$. In a single-electron picture, the $|f_\mathbf{a}\rangle$ state would have an electron in a state above Fermi level and a hole in some of its core states, while other electrons would occupy their original states, eventually modified by presence of a core hole. 
According to Coene and van Dyck \cite{coene} and Dwyer \cite{dwyer}, the beam wavefunction after the inelastic event can be described as $|\phi_{z=z_a}^{f_\mathbf{a}}\rangle = -i\sigma\hat{V}^\text{proj}_{i_\mathbf{a} \to f_\mathbf{a}} |\psi_{z=z_a}^{i_\mathbf{a}}\rangle$, where $\hat{V}^\text{proj}_{i_\mathbf{a} \to f_\mathbf{a}}$ is the projected interaction potential operator

\begin{equation} \label{eq:vproj}
  \hat{V}^\text{proj}_{i_\mathbf{a} \to f_\mathbf{a}} = \int \mathrm{d}z e^{-iq_{\Delta E}(z-z_a)}\langle z| \otimes \langle f_\mathbf{a} | \hat{V} | i_\mathbf{a} \rangle \otimes |z\rangle
\end{equation}
which acts on the beam electron wavefunction. The operator $\hat{V}$ describes the Coulomb interaction of the beam electron with charges in the sample and $q_{\mathbf\Delta E}$ being a characteristic momentum transfer due to the energy loss $\Delta E$ given by
\begin{equation} \label{eq:qdeltae}
  q_{\Delta E} = k_{f,z}-k_{i,z} \approx -k_{i,z} \frac{\Delta E}{2E}.
\end{equation}

Here one implicitly uses the so called $z$-locality approximation \cite{dwyer,verbeeck} (sometimes also called \emph{projection approximation}), assuming that the excitation process happened sharply at the $z$-coordinate of the excited atom, despite that the excitation is delocalized also along $z$ direction.

Next, the individual wavefunctions $|\phi_{z=z_a}^{f_\mathbf{a}}\rangle$ for all $\mathbf{a}$ and $f_\mathbf{a}$ are elastically propagated to the exit surface of the sample, to obtain $|\phi_{z=t}^{f_\mathbf{a}}\rangle = \hat{U}(t,z_a) |\phi_{z=z_a}^{f_\mathbf{a}}\rangle$. In writing this one assumes that an inelastic event has happened only once during a passage of a fast electron through the crystal, which is reasonable because of the generally small cross sections of core-loss events. Note furthermore that the propagator $\hat{U}$ depends on the electron energy, i.e., acts differently on $|\psi^{i_\mathbf{a}}\rangle$, which has the kinetic energy $E=eV_\text{acc}$, and $|\phi^{f_\mathbf{a}}\rangle$, with the kinetic energy $E'=eV_\text{acc} - E_{f_\mathbf{a}} + E_{i_\mathbf{a}}$, where $\Vacc$ is the acceleration voltage.
 Having evaluated all the $|\phi^{f_\mathbf{a}}\rangle$, we obtain the density matrix of the outgoing inelastically scattered electrons as an incoherent sum \cite{dwyer} in the following form
\begin{equation} \label{eq:rho}
  \hat{\rho}_f = \sum_\mathbf{a} \sum_{f_\mathbf{a}} | \phi_{z=t}^{f_\mathbf{a}}\rangle \langle \phi_{z=t}^{f_\mathbf{a}} |
\end{equation}
If necessary, the individual wavefunctions are passed further through the objective and projector lens, which are typically described by a transfer function \cite{kirkland}, discussed further below.

Once the density matrix is available, one can calculate intensity of electrons scattered into some direction $\mathbf{K}=(k_x,k_y)$ as
\begin{equation} \label{eq:kpix}
  I(\mathbf{K}) = \mathrm{Tr}\big[\hat{\rho}_f |\mathbf{K}\rangle\langle\mathbf{K}| \big] = \sum_{\mathbf{a},f_\mathbf{a}} \left| \langle\mathbf{K}|\phi_{z=t}^{f_\mathbf{a}}\rangle \right|^2
\end{equation}
eventually intensity of electrons scattered into an arbitrary detector described by a detector aperture function $A(\mathbf{K})$ given by
\begin{equation}
  I_A = \int A(\mathbf{K}) I(\mathbf{K}) d\mathbf{K}
\end{equation}
where for example
\begin{equation}
  A_\beta(\mathbf{K}) = \left\{ \begin{array}{rl}
                                  1 & \text{if } |\mathbf{K}| < 2\pi\beta/\lambda \\
                                  0 & \text{otherwise}
                                \end{array} \right.
\end{equation}
would describe a detector with a collection semi-angle $\beta$, where $\lambda$ is the wavelength of outgoing electrons.

Throughout the manuscript we use the following notation convention for vectors: two-dimensional vectors defined within an $x,y$-plane are typeset using capital letters, e.g., $\mathbf{K}=(k_x,k_y)$, while three-dimensional vectors are typeset using small letters, e.g., $\mathbf{k}$, one can thus write $\mathbf{k} = (\mathbf{K},k_z)$. Note that a state, e.g., $|\phi\rangle$ or $|\mathbf{K}\rangle$, when expressed in real space, $\langle\mathbf{r}|\phi\rangle = \phi(\mathbf{r})$ or $\langle\mathbf{r}|\mathbf{K}\rangle$, is always a three-dimensional object, despite that it can be parametrized by a two-dimensional wave-vector $\mathbf{K}$ (the $k_z$ is here fixed by acceleration voltage and eventually by energy loss). However, in some cases we collapse the $z$-coordinate of such states by specifying a plane, e.g., $|\phi_{z=z_a}\rangle$ or simply $|\phi_{z_a}\rangle = \langle z_a | \phi\rangle$. Then $\langle \mathbf{R} | \phi_{z_a}\rangle \equiv \phi(x,y;z=z_a)$ becomes two-dimensional, a function of $x,y$ only. In this context, the state $|\mathbf{K}\rangle$ introduced in Eq.~\ref{eq:kpix} means more precisely $|\mathbf{K}_{z=t}\rangle$, i.e., the state $|\mathbf{K}\rangle$ at the exit surface of the sample. Such a shortcut in the notation will only be used for states defined \emph{outside} the sample.

The equations above summarize the inelastic multislice approach to calculations of the inelastic scattering cross-section and serve as the starting point of our considerations. The structure of the paper is the following. 
In Section~\ref{sec:recwave} we summarize the reciprocal wave approach for energy filtered diffraction (EFDIF), high-resolution TEM imaging (HRTEM) and projections on general basis. In Section~\ref{sec:eval} we discuss the $\mathbf{k}$-space summation approach, which allows to evaluate inelastic scattering cross-sections using the concept of reciprocal waves.
One striking advantage of the reciprocal wave approach is a seamless integration of $z$-nonlocal inelastic interaction. In  Section~\ref{sec:comp} we present computational results exemplifying the above reciprocity as well as the impact of the $z$-locality approximation.

\section{Reciprocal wave}\label{sec:recwave}

So far we have only used the language of forward-propagation methods. Now we will present the concept of a reciprocal wave \cite{kainuma}, i.e., a wave propagating ``backwards in space'' from the detector into and through the sample. This section doesn't present new material, which the reader couldn't find elsewhere in the literature. However, we believe it will be found useful for the discussion in following sections, presented here in a compact notation consistent with the rest of the text, a notation which might differ from other literature on this subject.

First we consider a reciprocal beam originating from a point on a detector in the diffraction plane corresponding to a single plane wave $|\mathbf{K}\rangle$ in the object plane. Second, we will consider a reciprocal wave originating from a point on a detector set to acquire an image in real space, essentially corresponding to the point spread function of the detection system in the object plane. In the last subsection we will briefly generalize the concept to other basis sets of possible interest.

\subsection{Energy Filtered Diffraction Pattern}

We rewrite Eq.~\ref{eq:kpix} 
from the Introduction section
\begin{eqnarray}
  I(\mathbf{K}) & = & \text{Tr} \big[ |\mathbf{K}\rangle\langle\mathbf{K}| \hat{\rho}_f \big] = \nonumber \\
                & = & \sum_{\mathbf{a},f_\mathbf{a}} \text{Tr} \big[ |\mathbf{K}\rangle\langle\mathbf{K}|\phi_{t}^{f_a}\rangle\langle\phi_{t}^{f_a}| \big] \nonumber \\
                & = & \sum_{\mathbf{a},f_\mathbf{a}} \text{Tr} \big[ |\mathbf{K}\rangle\langle\mathbf{K}|\hat{U}(t,z_a)|\phi_{z_a}^{f_a}\rangle\langle\phi_{z_a}^{f_a}|\hat{U}^\dagger(t,z_a) \big] \nonumber \\
                & = & \sum_{\mathbf{a},f_\mathbf{a}} \langle\mathbf{K}|\hat{U}(t,z_a)|\phi_{z_a}^{f_a}\rangle\langle\phi_{z_a}^{f_a}|\hat{U}^\dagger(t,z_a)|\mathbf{K}\rangle \nonumber \\
                & = & \sum_{\mathbf{a},f_\mathbf{a}} \langle\mathbf{K}|\hat{U}^\dagger(z_a,t)|\phi_{z_a}^{f_a}\rangle\langle\phi_{z_a}^{f_a}|\hat{U}(z_a,t)|\mathbf{K}\rangle ,
\end{eqnarray}
where we have utilized the unitarity of the evolution operator. Now we realize that $\hat{U}(z_a,t)|\mathbf{K}\rangle$ is just a plane wave $\mathbf{K}$ propagated ``back in time'' into the crystal  (back in $z$-coordinate, actually). That is exactly the concept of a reciprocal (backpropagated) wave: a wave entering the crystal at the exit surface and propagating in a direction opposite to the beam. 

If we denote its state at coordinate $z_a$ as
\begin{equation}
   |\mathbf{K}_{z_a}^{BP}\rangle \equiv \hat{U}(z_a,t)|\mathbf{K}\rangle,
\end{equation}
then the resulting pixel of an EFDIF pattern can be written as
\begin{equation}
  I(\mathbf{K}) = \sum_{\mathbf{a},f_\mathbf{a}} | \langle \mathbf{K}_{z_a}^{BP} | \phi_{z_a}^{f_a} \rangle |^2 .
\end{equation}

Focusing now on the incoming wave, first elastically propagated and then scattered inelastically, we carry out the substitution mentioned in the Introduction section
\begin{equation}
  | \phi_{z_a}^{f_a} \rangle = -i \sigma \hat{V}^\text{proj}_{i_a \to f_a} | \psi^{i_a}_{z_a} \rangle
\end{equation}
to obtain
\begin{equation} \label{eq:iK}
  I(\mathbf{K}) = \sum_{\mathbf{a},f_\mathbf{a}} | \langle \mathbf{K}_{z_a}^{BP} | \sigma \hat{V}^\text{proj}_{i_a \to f_a} | \psi_{z_a} \rangle |^2 .
\end{equation}

\subsection{Energy-filtered HRTEM images}

Instead of calculating a diffraction pattern, as represented by $\mathrm{Tr}\big[ \hat{\rho} |\mathbf{K}\rangle\langle\mathbf{K}| \big]$, we may aim at computing a high-resolution TEM image at a specific energy loss, i.e., an EF-HRTEM image, $\mathrm{Tr}\big[ \hat{\rho} |\mathbf{R}\rangle\langle\mathbf{R}| \big]$, where $\mathbf{R}$ labels a detector coordinate associated with the $(x,y,t)$ coordinate in the object plane. For simplicity, we will not discuss issues of partial coherence here and assume a fully coherent imaging.

In EF-HRTEM imaging the wave passes through an imaging lens, potentially including an image aberration corrector, and a limiting aperture, which only lets pass the electrons that have been scattered below a certain maximum angle. The whole optical transfer is typically described by a transmission function $T(\mathbf{K})$. Accordingly, an observed image is given by
\begin{equation}
  I(\mathbf{R}) = \int \!\! d\mathbf{K} \int \!\! d\mathbf{K'} \mathrm{Tr} \big[ |\mathbf{R}\rangle\langle\mathbf{R}|\mathbf{K}\rangle T^\star(\mathbf{K}) \langle\mathbf{K}| \hat{\rho}  |\mathbf{K'}\rangle T(\mathbf{K'}) \langle \mathbf{K'} | \big] .
\end{equation}
The expression
\begin{equation}
  \hat{T} = \int d\mathbf{K} |\mathbf{K}\rangle T(\mathbf{K}) \langle \mathbf{K} |
\end{equation}
can be also understood as a projection operator on a convergent reciprocal wave (with convergence angle equal to the collection angle of the limiting aperture) with a phase distribution defined by the aberrations of the optics, as contained in the transmission function $T(\mathbf{K})$. Furthermore, the factor $\langle \mathbf{R} | \mathbf{K} \rangle = e^{i\mathbf{K}\cdot\mathbf{R}}$ can be understood as a phase ramp for the reciprocal wave. According to the Fourier shift theorem, this ramp originates from shifting the ``point source'' (point on the detector) of the convergent reciprocal wave.

Similarly as in the previous section, we can construct a reciprocal wave of the form
\begin{equation}
  \hat{T} | \mathbf{R} \rangle = \int d\mathbf{K} |\mathbf{K}\rangle T(\mathbf{K}) \langle \mathbf{K} | \mathbf{R} \rangle ,
\end{equation}
which we back-propagate into the crystal. We denote such a bra vector by
\begin{equation}
  \langle \mathbf{R}^{BP}_{T,z_a}| = \int d\mathbf{K} \langle \mathbf{R} |\mathbf{K}\rangle T^\star(\mathbf{K}) \langle \mathbf{K} | \hat{U}(t,z_a) ,
\end{equation}
where we assume the evolution operator acting at a specific value of the kinetic energy. Finally, the image intensity can be written as
\begin{equation} \label{eq:iR}
  I(\mathbf{R}) = \sum_{\mathbf{a},f_\mathbf{a}} | \langle \mathbf{R}^{BP}_{T,z_a} | \sigma \hat{V}^\text{proj}_{i_a \to f_a} | \psi_{z_a} \rangle  |^2 .
\end{equation}

\subsection{Other bases}

The reciprocal plane waves represent only one particular basis set. Any outgoing wave can be expanded into plane waves, but equally so we could have chosen a different basis set. For example, if the outgoing wave would be filtered by its orbital angular momentum (OAM) character, e.g., let's say the apertures would let pass only the OAM=1$\hbar$ part of the outgoing wave, we could calculate its intensity by a corresponding projection of the density matrix to associated basis functions. Here we could use for example Laguerre-Gauss modes \cite{allenEVB,bliokhLG} $L_p^l$, where $l$ denotes the angular momentum, and $p$ labels the radial part of the wavefunction:
\begin{equation}
  I(l,p) = \text{Tr} \big[ \hat{\rho} |L_l^p\rangle \langle L_l^p| \big] .
\end{equation}
One would need to calculate $I(l=1,p)$ for a sufficient range of $p$ values and their sum would represent the intensity of the outgoing beam of OAM=1$\hbar$ character. 

In general, it is important that the basis is complete in the two-dimensional space of the outgoing wavefunctions (one dimension is fixed by the energy of the outgoing beam). It does not matter, whether it is parametrized by $(k_x,k_y)$, $(R_x,R_y)$ or $(l,p)$ or yet other parameters, like for example $(l,k_\perp)$ for Bessel beams. Such basis functions, let's denote them $|\phi_{u,v}\rangle$, parametrized by $u,v$, need to be back-propagated to obtain $\langle \phi_{u,v}^{BP}|$ as was described in the previous subsection, and then enter the summation as the outgoing wavefunction instead of $\langle\mathbf{K}^{BP}|$ in Eq.~\ref{eq:iK} or $\langle \mathbf{R}^{BP}_{T}|$ in Eq.~\ref{eq:iR}, respectively:
\begin{equation} \label{eq:Iuv}
  I(u,v) = \sum_{\mathbf{a},f_\mathbf{a}} | \langle \phi^{BP}_{u,v;z_a} | \sigma \hat{V}^\text{proj}_{i_a \to f_a} | \psi_{z_a} \rangle  |^2 .
\end{equation}

\section{K-space summation and fully non-local calculations}\label{sec:eval}

In the previous section we have presented a formal manipulation of the scattering cross-section formula using the concept of reciprocal waves. The $z$-locality approximation was present throughout the whole section. Here we show, how we can incorporate $z$-nonlocality. In the second subsection we show how this can be evaluated using the $\mathbf{k}$-space summation methods\cite{rossouw,kohl,saldin,schattbw,prbtheory} and finally, we reintroduce the $z$-locality approximation within the $\mathbf{k}$-space summation formalism, to have a computational method allowing to easily switch the $z$-locality approximation on and off.

\subsection{Abandoning $z$-locality}
\,
To abandon $z$-locality we can apply the first Born approximation: instead of using a projected transition potential operator $\hat{V}^\text{proj}_{i_a \to f_a}$ we use its non-projected counterpart
\begin{equation}
  \hat{V}_{i_a \to f_a} = \langle f_a | \hat{V} | i_a \rangle ,
\end{equation}
and instead of evaluating two-dimensional integrals in Eq.~\ref{eq:Iuv} we evaluate three-dimensional integrals over the incoming and outgoing beam wavefunctions:
\begin{equation} \label{eq:Iuvnonloc}
  I(u,v) = \sum_{\mathbf{a},f_\mathbf{a}} | \langle \phi_{u,v}^{BP} | \sigma \hat{V}_{i_a \to f_a} | \psi \rangle |^2 .
\end{equation}
Note the formal similarity of Eqns.~\ref{eq:Iuv} and \ref{eq:Iuvnonloc}. However, their evaluation is rather different. When the matrix elements are expressed in real space, Eq.~\ref{eq:Iuv} is a sum of two-dimensional integrals, while Eq.~\ref{eq:Iuvnonloc} is a sum of three-dimensional integrals. In the next subsection we will summarize, how this expression can be evaluated by means of $\mathbf{k}$-space summation.


\subsection{K-space summation}

By expanding Eq.~\ref{eq:Iuvnonloc} in reciprocal space, we obtain
\begin{eqnarray}
  I(u,v) & = & \sum_{\mathbf{a},f_\mathbf{a}} \Big| \int d\mathbf{k'} \int d\mathbf{k''} \nonumber \\
                & \times & \langle \phi_{u,v}^{BP} | \mathbf{k'} \rangle \langle \mathbf{k'} | \sigma \hat{V}_{i_a \to f_a} | \mathbf{k''} \rangle \langle \mathbf{k''} | \psi \rangle \Big|^2 .
\end{eqnarray}
Focusing on a narrow range of energy losses we select only those $f_a$, which fulfill $E_{f_a}-E_{i_a} = \Delta E$ for a fixed $\Delta E$.
Introducing the notation
\begin{equation} \label{eq:bcoef}
  D_\mathbf{k'} = \langle \mathbf{k'} | \phi_{u,v}^{BP} \rangle \qquad \text{and} \qquad C_\mathbf{k''} = \langle \mathbf{k''} | \psi \rangle
\end{equation}
we can rewrite the expression above as
\begin{equation}
  I(u,v) = \sigma^2 \sum_{\mathbf{a},f_\mathbf{a}} \Big| \int \!\! d\mathbf{k'} \int \!\! d\mathbf{k''} D^\star_\mathbf{k'} C_\mathbf{k''} \langle \mathbf{k'},f_a | \hat{V} | \mathbf{k''},i_a \rangle \Big|^2
\end{equation}
where the star marks complex conjugation. After expanding the square, changing the order of integration and summation, and using
\begin{equation}\label{eq:ftcoul}
  \langle \mathbf{k'},f_a | \hat{V} | \mathbf{k''},i_a \rangle = \frac{4\pi}{|\mathbf{k'-k''}|^2} \langle f_a | e^{i\mathbf{(k'-k'')\cdot r}} | i_a \rangle,
\end{equation}we finally obtain
\begin{eqnarray}\label{eq:ikksum}
  I(u,v) & = & \sigma^2 \int \ldots \int d\mathbf{k}_1 \ldots d\mathbf{k}_4 \nonumber \\
                & \times & D^\star_{\mathbf{k}_1} C_{\mathbf{k}_2} D_{\mathbf{k}_3} C^\star_{\mathbf{k}_4} \frac{S(\mathbf{q,q'},\Delta E)}{q^2 q'^2} ,
\end{eqnarray}
with
\begin{eqnarray}
  \mathbf{q} & = & \mathbf{k}_1 - \mathbf{k}_2 \label{eq:qvec} \\
  \mathbf{q'} & = & \mathbf{k}_3 - \mathbf{k}_4 \label{eq:qprvec}
\end{eqnarray}
and the mixed dynamic form factor \cite{kohl} (MDFF) of the sample
\begin{eqnarray}\label{eq:mdff}
  S(\mathbf{q,q'},\Delta E) & = & \sum_{\mathbf{a},f_\mathbf{a}} \langle i_a | e^{i\mathbf{q'\cdot r}} | f_a \rangle \langle f_a | e^{-i\mathbf{q\cdot r}} | i_a \rangle \nonumber \\
                            & \times & \delta(E_{f_a}-E_{i_a}- \Delta E)
\end{eqnarray}
This total MDFF can be written as a sum of atomic MDFFs, $S_\mathbf{a}(\mathbf{q,q'},\Delta E)$, multiplied by phase factors
\begin{equation} \label{eq:mdfftot}
  S(\mathbf{q,q'},\Delta E) = \sum_\mathbf{a} e^{i\mathbf{(q'-q)\cdot a}} S_\mathbf{a}(\mathbf{q,q'},\Delta E),
\end{equation}
where the atomic MDFF
\begin{eqnarray}\label{eq:mdffat}
  S_\mathbf{a}(\mathbf{q,q'},\Delta E) & = & \sum_{f_\mathbf{a}} \langle i_\mathbf{a} | e^{i\mathbf{q'\cdot (r-a)}} | f_\mathbf{a} \rangle \langle f_\mathbf{a} | e^{-i\mathbf{q\cdot (r-a)}} | i_\mathbf{a} \rangle \nonumber \\
                            & \times & \delta(E_{f_\mathbf{a}}-E_{i_\mathbf{a}}- \Delta E)
\end{eqnarray}
can be efficiently evaluated in a local coordinate system centered on atom $\mathbf{a}$\cite{kohl,rossouw,saldin,schattbw,prbtheory}. Note that the atomic MDFF can be formally expressed in dipole approximation as\cite{opmaps}
\begin{equation}
  S_\mathbf{a}(\mathbf{q,q'},E) = \mathbf{q}\cdot\mathbb{N}_\mathbf{a}(E)\cdot\mathbf{q'} + i(\mathbf{q \times q'})\cdot\mathbf{M}_\mathbf{a}(E)
\end{equation}
where $\mathbb{N}_\mathbf{a}(E)$ is a real-valued symmetric tensor containig information about the non-magnetic part of the electronic structure and $\mathbf{M}_\mathbf{a}(E)$ is a vector containing the information about magnetism \cite{nature} of atom $\mathbf{a}$. This allows to factor out the electronic structure information from the dynamical diffraction calculation and to calculate normalized images, e.g., per hole in the $d$-shell, or per $1\mu_B$ of spin magnetization in $z$-direction, etc. Moreover, it is possible to use more precise approximations, including monopole, quadrupole, octupole transitions, etc., including their cross-terms, e.g., by employing more efficient spherical Bessel function expansions \cite{loffler}.

Equation~\ref{eq:Iuvnonloc} and the subsequent derivation leading to Eq.~\ref{eq:ikksum} shows, how the two approaches for calculating inelastic scattering cross-section, the reciprocal wave and the forward integration approach, can be related to each other. One main difference is the $z$-locality of the inelastic event assumed in the forward propagation based approaches. The second one pertains to the organization of the computations itself. In the reciprocal wave approach, we don't have to start a new independent propagation for each possible inelastic transition at every atom in the sample. Instead, the $I(u,v)$ needs to be evaluated for all needed combinations of parameters $u,v$. For instance, if they represent $(u,v) \equiv (k_x,k_y) = \mathbf{K}$, then $I(\mathbf{K})$ needs to be evaluated for each pixel on the desired grid of EFDIF pattern. In some situations this can save large amounts of computing time, particularly when the $\mathbf{k}$-space summation in Eq.~\ref{eq:ikksum} is implemented in an efficient way \cite{bwconv,bwconv2}. The price paid is that we do not know the individual exit wavefunctions, or more precisely, the exit density matrix. Instead we are directly obtaining the scattering cross-section.

Equation~\ref{eq:ikksum} resembles the Bloch waves method of calculation of the inelastic scattering cross-section\cite{rossouw,saldin,kohl,schattbw,prbtheory}, it is however more general. The incoming beam and backpropagated beam wavefunctions can be calculated by any method, e.g., Bloch waves or multislice method\cite{cowley,kirkland}, as long as we can expand these wavefunctions in $\mathbf{k}$-space to obtain Fourier coefficients $C_\mathbf{k},D_\mathbf{k'}$. For instance, the \textsc{mats}\cite{vortexsurvey} and \textsc{mats.v2}\cite{bwconv2} algorithms use Bloch waves for the backpropagated wavefunction and conventional multislice for the incoming wavefunction.

\subsection{Re-introducing the z-locality approximation}\label{sec:zloc}

Let's explicitly evaluate Eq.~\ref{eq:Iuv} using the $\mathbf{k}$-space summation and compare it to the fully non-local expressions from the previous subsection. The projected potential is given by Eq.~\ref{eq:vproj} and the two-dimensional slices of a general wavefunction $|\psi\rangle$ that we denoted $|\psi_{z_a}\rangle$ is given by $\langle z_a | \psi \rangle$. The reader should be aware that this is still a state in a two-dimensional Hilbert space, not a scalar. We can thus expand a non-local term from Eq.~\ref{eq:Iuv} in the following way:
\begin{eqnarray} \label{eq:termzloc}
  \lefteqn{\langle \phi^{BP}_{u,v;z_a} | \hat{V}^\text{proj}_{i_a \to f_a} | \psi_{z_a} \rangle =  } \nonumber \\
  & = & \langle \phi_{u,v}^{BP} | z_a\rangle \int \mathrm{d}z e^{-iq_{\Delta E}(z-z_a)}\langle z| \otimes \langle f_\mathbf{a} | \hat{V} | i_\mathbf{a} \rangle \otimes |z\rangle \langle z_a | \psi \rangle \nonumber \\ 
  & = & \iiint \!\! \mathrm{d}\mathbf{k} \mathrm{d}\mathbf{k'} \mathrm{d}z \langle \phi_{u,v}^{BP} | \mathbf{k'} \rangle \langle \mathbf{k'} | z_a\rangle  \nonumber \\
  & \times &  e^{i q_{\Delta E} z_a} e^{-i(k_{f,z}-k_{i,z})z}\langle z| \otimes \langle f_\mathbf{a} | \hat{V} | i_\mathbf{a} \rangle \otimes |z\rangle \nonumber \\
  & \times & \langle z_a | \mathbf{k}\rangle \langle \mathbf{k}| \psi \rangle \nonumber \\
  & = & \iiint \!\! \mathrm{d}\mathbf{k} \mathrm{d}\mathbf{k'} \mathrm{d}z D_\mathbf{k'}^\star e^{-ik'_z z_a} C_\mathbf{k} e^{ik_z z_a} e^{iq_{\Delta E}z_a} \nonumber \\
  & \times &  e^{-ik_{f,z}z} \langle z| \langle \mathbf{K'} | \otimes \langle f_\mathbf{a} | \hat{V} | i_\mathbf{a} \rangle \otimes |\mathbf{K}\rangle |z\rangle e^{ik_{i,z}z} \nonumber \\
  & = & \iint \!\! \mathrm{d}\mathbf{k} \mathrm{d}\mathbf{k'} D_\mathbf{k'}^\star C_\mathbf{k} e^{i(q_{\Delta E}+k_z-k'_z)z_a} \langle\mathbf{\tilde{k}'},f_\mathbf{a}|\hat{V} | i_\mathbf{a},\mathbf{\tilde{k}}\rangle \nonumber \\
  & = & \iint \!\! \mathrm{d}\mathbf{k} \mathrm{d}\mathbf{k'} D_\mathbf{k'}^\star C_\mathbf{k} e^{i(q_{\Delta E}+k_z-k'_z)z_a} \frac{4\pi \langle f_\mathbf{a}| e^{i(\mathbf{\tilde{k}'-\tilde{k}})\cdot\mathbf{r}} |i_\mathbf{a} \rangle}{|\mathbf{\tilde{k}'-\tilde{k}}|^2}
\end{eqnarray}
where we used Eqns.~\ref{eq:qdeltae}, \ref{eq:bcoef} and \ref{eq:ftcoul}, relation $\langle z_a | \mathbf{k}\rangle = e^{ik_z z_a}|\mathbf{K}\rangle$, and introduced notation $\mathbf{\tilde{k}}=(\mathbf{K},k_{i,z})$ and $\mathbf{\tilde{k}'}=(\mathbf{K'},k_{f,z})$.

The difference from a fully non-local expression consists thus of a phase factor $e^{i(q_{\Delta E}+k_z-k'_z)z_a}$ and replacement of $\mathbf{k,k'}$ in the Fourier transformed matrix element of the Coulomb interaction by $\mathbf{\tilde{k},\tilde{k}'}$. Taking a sum over $\mathbf{a},f_\mathbf{a}$ of the squared absolute value of Eq.~\ref{eq:termzloc} we obtain an expression identical to Eq.~\ref{eq:ikksum}
\begin{eqnarray}\label{eq:ikksum2}
  I(u,v) & = & \sigma^2 \int \ldots \int d\mathbf{k}_1 \ldots d\mathbf{k}_4 \nonumber \\
                & \times & D^\star_{\mathbf{k}_1} C_{\mathbf{k}_2} D_{\mathbf{k}_3} C^\star_{\mathbf{k}_4} \frac{S(\mathbf{\tilde{q},\tilde{q}'},\Delta E)}{\tilde{q}^2 \tilde{q}'^2} ,
\end{eqnarray}
if we redefine $\mathbf{q,q'}$ using 
\begin{eqnarray}
  \mathbf{\tilde{q}}  & = & \mathbf{K}_1 - \mathbf{K}_2 + q_{\Delta E}\mathbf{\hat{z}} \label{eq:qtilde} \\
  \mathbf{\tilde{q}'} & = & \mathbf{K}_3 - \mathbf{K}_4 + q_{\Delta E}\mathbf{\hat{z}} \label{eq:qprtilde}
\end{eqnarray}
and absorb the phase factor $e^{-i(k_{4,z}-k_{3,z})z_a}e^{i(k_{2,z}-k_{1,z})z_a}$ into the total MDFF. Expressing then the total MDFF as a sum of atomic MDFFs, including this phase factor, leads to
\begin{eqnarray}
  S(\mathbf{\tilde{q},\tilde{q}'},\Delta E) & = & e^{i[(k_{3,z}-k_{4,z})-(k_{1,z}-k_{2,z})]z_a} \nonumber \\
  & \times & \sum_\mathbf{a} e^{i\mathbf{(\tilde{q}'-\tilde{q})\cdot a}} S_\mathbf{a}(\mathbf{\tilde{q},\tilde{q}'},\Delta E) \nonumber \\
  & = & \sum_\mathbf{a} e^{i\mathbf{(q'-q)\cdot a}} S_\mathbf{a}(\mathbf{\mathbf{\tilde{q}},\mathbf{\tilde{q}}'},\Delta E)
\end{eqnarray}
Therefore the $z$-locality approximation within the $\mathbf{k}$-space summation approach is achieved merely by replacing $\mathbf{q,q'}$ by $\mathbf{\tilde{q},\tilde{q}'}$ (see Eqns.~\ref{eq:qtilde} and \ref{eq:qprtilde}) in the atomic MDFFs and their associated Coulomb factors $1/\tilde{q}^2\tilde{q}'^2$, while all the other prefactors remain the same as in the fully non-local treatment. It is thus trivial to switch between the $z$-local and fully non-local calculations in simulations based on $\mathbf{k}$-space summation.

\subsection{Implementation note about EFTEM simulations}\label{sec:invert}

In the EF-HRTEM case the incoming beam is typically a single plane wave entering the sample in a direction perpendicular to the surface. While from the formal point of view, this has no bearing for the argumentation until this point, for an actual realization of calculations it is a very useful observation, because it allows us to use existing codes originally dedicated to EFDIF or SI calculations \cite{bwconv2} with minimal modifications. We only have to ``invert the microscope'', which is formally trivial. Again, we silently assume that we focus on energy loss processes in a narrow energy range around $\Delta E$:
\begin{eqnarray}
  I(\mathbf{R}) & = & \sum_a | \langle \mathbf{R}^{BP}_T | \sigma \hat{V}_{i_a \to f_a} | \psi \rangle  |^2 \nonumber \\
  & = & \sum_a \sigma^2 | \langle \mathbf{R}^{BP}_T | \langle f_a | \hat{V} | i_a \rangle | \psi \rangle  |^2 \nonumber \\
  & = & \sum_a \sigma^2 | \langle \psi | \langle i_a | \hat{V} | f_a \rangle | \mathbf{R}^{BP}_T \rangle  |^2 \nonumber \\
  & = & \sum_a | \langle \psi | \sigma \hat{V}_{f_a \to i_a} | \mathbf{R}^{BP}_T \rangle  |^2
\end{eqnarray}
Here, we have used again that the interaction potential is a hermitean operator. The last expression is nothing else than an elastically scattered convergent electron beam of kinetic energy $eV_\text{acc} - \Delta E$ going through an energy gain process, projected on a back-propagated reciprocal plane wave. Within this picture, the collection angle becomes the convergence angle and aberrations of the projector optics turn into aberrations of the probe forming optics. Due to hermitean conjugation of the transmission operator $\hat{T}$, the actual aberrations of projector optics should be multiplied by a factor $-1$, if we aim to treat this process as an energy gain process of a convergent probe. By using this trick we can readily compute EFTEM imaging with the $\textsc{mats}$ or $\textsc{mats.v2}$ algorithm, as long as we treat the incoming beam as having kinetic energy $eV_\text{acc}-\Delta E$ and the outgoing beam having kinetic energy $eV_\text{acc}$. Accordingly, the energy loss becomes negative $\Delta E \to -\Delta E$.


Note that this statement differs from the reciprocity theorem\cite{recipth}, which states: \emph{``The amplitude at B of a wave originating from a source at A, and scattered by P, is equal to the scattered amplitude at A due to the same source placed at B.''} In our case, in the reciprocal process, we are sending from $B$ a completely different wave than what originated from the source $A$ in the original process. 
Yet the relation between the cross-sections holds, because it is essentially formulated as a square of a transition matrix element of a hermitean interaction potential operator, and that allows us to swap the incoming and outgoing waves and direction of their propagation and energy loss processes. Note also that this relation holds without any additional approximations needed for the reciprocity theorem, when considering inelastic processes\cite{recipth} ($k_{f_\mathbf{a}} \approx k_{i_\mathbf{a}}$). 

Once the equivalence of the scattering cross-sections has been established, we can proceed with simulating the EF-HRTEM images using $\mathbf{k}$-space summation methods developed for STEM-SI\cite{bwconv,bwconv2,recipschat}.

\section{Computational Case Studies}\label{sec:comp}

In the first subsection we compare EF-HRTEM simulations of SrTiO$_3$ using the modified $\textsc{mats.v2}$ algorithm to published results and also address the impact of $z$-locality approximation on the calculated images. In the second subsection we discuss impact of the $z$-locality approximation on EF-HRTEM images of PbZrO$_3$ as a function of core-level edge, thickness and defocus.

\subsection{Strontium titanate}\label{sec:sto110}

\begin{figure*}[th]
  \includegraphics[width=17.9cm]{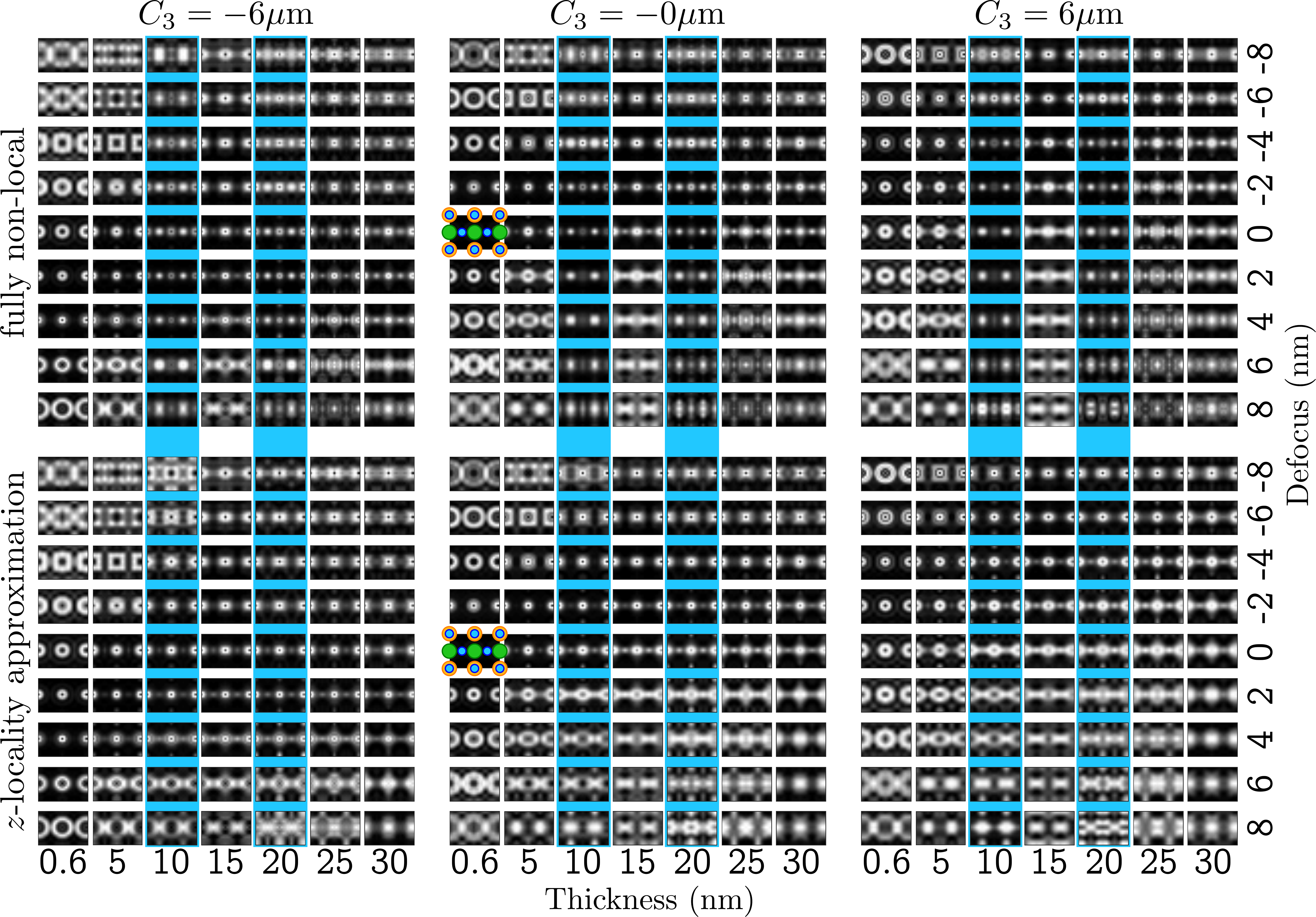}
  \caption{Simulations of Ti $L_3$ edge energy-filtered high-resolution TEM images of SrTiO$_3$ in [110] zone axis orientation for a range of defoci and spherical aberrations. The upper row represents simulations with full non-locality of the inelastic scattering, while in the bottom we have applied the $z$-locality approximation (similar to Forbes at al.\cite{forbessto}). The position of atomic columns is marked in panels with zero defocus and spherical aberration, at thickness 0.6~nm. Ti columns are marked in green color, Sr columns in orange color, and O columns are marked with smaller spheres of blue color. Dimensions of each individual plot are 5.52~\AA{} $\times$ 3.91~\AA{}. Blue frames mark thicknesses, where the largest differences between fully non-local and $z$-local treatment can be observed, see text for more details.}
  \label{fig:sto110}
\end{figure*}

In this section we perform simulations of Ti $L_3$ edge ($\Delta E=456$~eV) EF-HRTEM images of SrTiO$_3$ oriented along the [110] zone axis. This system has been analyzed in detail by Forbes et al.\cite{forbessto} previously. They employed multislice computations with the $z$-locality approximation, and hence provide a reference for our reciprocal wave computations. There are some subtle differences in settings of our calculations, e.g., in the Debye-Waller factors, the initial and final atomic wave functions (we use a simple dipole approximation), and the smooth envelope defining the detector\cite{forbessto}, replaced in our case by a sharp circular aperture with collection semi-angle of 37.5~mrad, which corresponds to the information limit quoted by Forbes et al.

For our calculations we have prepared an orthogonal supercell of SrTiO$_3$ containing two formula units of SrTiO$_3$, which has $c$-axis parallel to the $[110]$ direction of the primitive unit cell. Lattice parameters of the simple cubic unit cell are $a=3.905$~\AA{} and the supercell has lattice parameters $\sqrt{2}a \times a \times \sqrt{2}a$. Calculations of EF-HRTEM images were performed for 7 thicknesses, comprising approximately 0.6, 5, 10, 15, 20, 25 and 30~nm, corresponding to Fig.~8 in Forbes et al.\cite{forbessto}. The acceleration voltage was set to 200~kV, the $C_5$ aberration was set to 1.5~mm and the defoci and spherical aberrations were varied from -8~nm to 8~nm, and -6~$\mu$m to 6~$\mu$m, respectively. All other aberrations were set to zero, following the above-mentioned reference.

We have performed calculations with full non-locality in $z$-direction, as well as with the $z$-locality approximation, as introduced in Sec.~\ref{sec:zloc}. Results of our simulations are summarized in Fig.~\ref{fig:sto110}.

First of all, our simulations with $z$-locality approximation are in rather tight agreement with the results of Forbes et al., although some small differences can be spotted. We attribute them to the technical differences discussed above. Yet the qualitative features and trends are matching across the whole range of the considered parameters space, thus yielding satisfactory agreement. A posteriori, we consider this to be a strong validation check of our computational approach, in particular confirming the equivalence between forward scattering and reciprocal wave approach within the $z$-locality approximation.

The most intriguing findings originate, however, from the comparison of the calculations with $z$-locality to the calculations with full non-locality of the inelastic transition. When inspecting the $C_3=0$ results, we mainly observe an underestimation of the dechanneling into the oxygen columns when employing the $z$-locality approximation. Curiously, however, at thicknesses 5, 15 and to a lesser extent 25~nm the impact of the $z$-locality approximation appears to be relatively minor, independent from the defocus. Contrary to that, at 10, 20 and to a lesser extent 30~nm the two calculations differ significantly from each other. Moreover, the fully non-local calculation tends to show somewhat sharper features in the EF-HRTEM images.

Both, the increased dechanneling and the sharper features in EF-HRTEM images, when considering fully non-local inelastic interactions, could be traced back to a systematic underestimation of large angle scattering in the $z$-locality approximation. 
To explain this effect, we note again that both the incoming and back-propagated electron beam wavefunctions enter the two simulations (local and non-local) in the exactly same form. Mathematically the only difference stems from approximated MDFFs and the associated Coulomb factors (Sec.~\ref{sec:zloc}), which serve as weights for individual products of the Fourier components of the electron beam wavefunction. Thus, the enhanced large angle scattering within the fully-delocalized picture may be attributed the subset of inelastic transitions involving terms $\frac{S(\mathbf{q,q'},\Delta E)}{q^2q'^2}$ with very small denominators. It is most convenient to illustrate that using the Bloch wave picture \cite{kainuma,saldin,prbtheory}. Within the Bloch wave method, the elastically scattered incoming wavefunction (or reciprocal backpropagated wavefunction) is expanded into a coherent superposition of Bloch states of a specific energy. Such Bloch states are characterized by wavevectors $\mathbf{k}^{(j)}$ with $k_z^{(j)} = k_z + \gamma^{(j)}$, where $\gamma^{(j)}$ is the so called elongation (also: \emph{Anpassung}) of the wavevector. There are separate sets of wavevector elongations $\gamma^{(j)}, \gamma^{(l)}$ for incoming and outgoing electron beam wavefunction and it can happen that they can approximately compensate the difference $k_{f,z}-k_{i,z}$. In such case the momentum transfer vectors can become arbitrarily small, which cannot happen within the $z$-locality approximation, because there the minimal magnitude of a momentum transfer vector is $q_{\Delta E}$. That can significantly enhance the weight of some terms in the summation of DDSCS.

\begin{figure}
  \includegraphics[width=8.6cm]{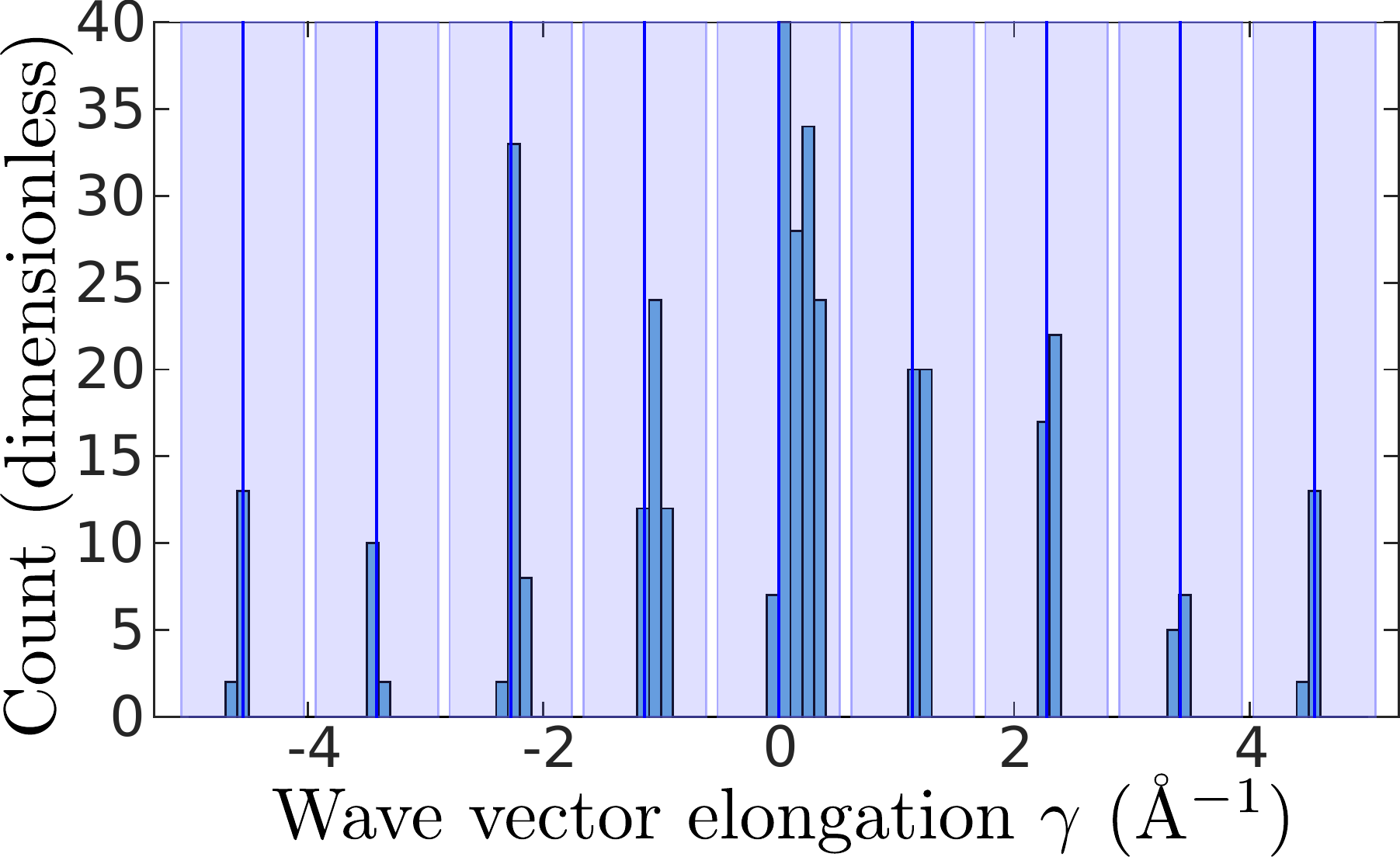}
  \caption{Histogram of wavevector elongations $\gamma$ for a plane wave of energy 200~keV, incoming along the $[110]$ direction, scattering on SrTiO$_3$ crystal. Vertical blue solid lines mark the multiples of $c^\star$ axis, the reciprocal lattice vector of SrTiO$_3$ [110] supercell, and the light blue shaded areas around them denote the $\pm q_{\Delta E}$ region. Note that in the first Brillouin zone there are plenty of $\gamma$'s that are in size comparable to $q_{\Delta E}$.}
  \label{fig:gammas}
\end{figure}

Note that this is not in contradiction with the argumentation justifying the $z$-locality approximation found in the Appendix of Verbeeck et al.\cite{verbeeck}, except for the assumption that $q_{\Delta E} \gg \gamma^{(j,l)}$. Fig.~\ref{fig:gammas} shows a histogram of $\gamma^{(j)}$ values for an incoming plane wave beam along the $z$-direction used in the SrTiO$_3$ calculations. Note that there is a number of individual $\gamma$-values that are comparable in magnitude to $q_{\Delta E}$ and therefore a situation, in which a combination $\gamma^{(j)}-\gamma^{(l)}$ is approximately equal to $q_{\Delta E}$, is likely to happen. In fact, this should not be surprising. Multislice simulations of elastic scattering have shown that the beam wavefunction shows short-wavelength ripples in its amplitude, particularly in the close neighborhood of atoms\cite{axel} in addition to the long-wavelength \emph{Pendell\"{o}sung} oscillations. The former ones are necessarily connected with sufficiently long $\gamma$-elongations from the Bloch waves perspective.

\begin{figure}
	\includegraphics[]{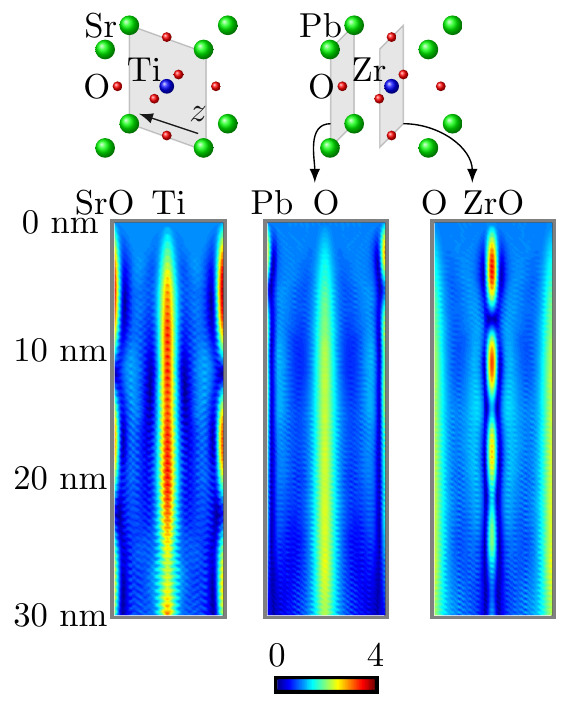}
	\caption{Periodic channeling of the wave $\psi^i$ elastically scattered on SrTiO$_3$ and PbZrO$_3$. In SrTiO$_3$ the wavelength of the channeling amounts to approximately 11 nm at the SrO column, 27 nm at the Ti column. Note the suboscillation with maxima at approximately 10 and 19 nm at the Ti column. In PbZrO$_3$ periodic channeling (with a wavelength of approximately 7 nm) is only observed at the ZrO column, whereas the scattering power of the Pb column quickly disperses the focussing effect beyond the first maxima.
	}
	\label{fig:channeling}
\end{figure}

\begin{figure*}[thb]
  \includegraphics[width=17.9cm]{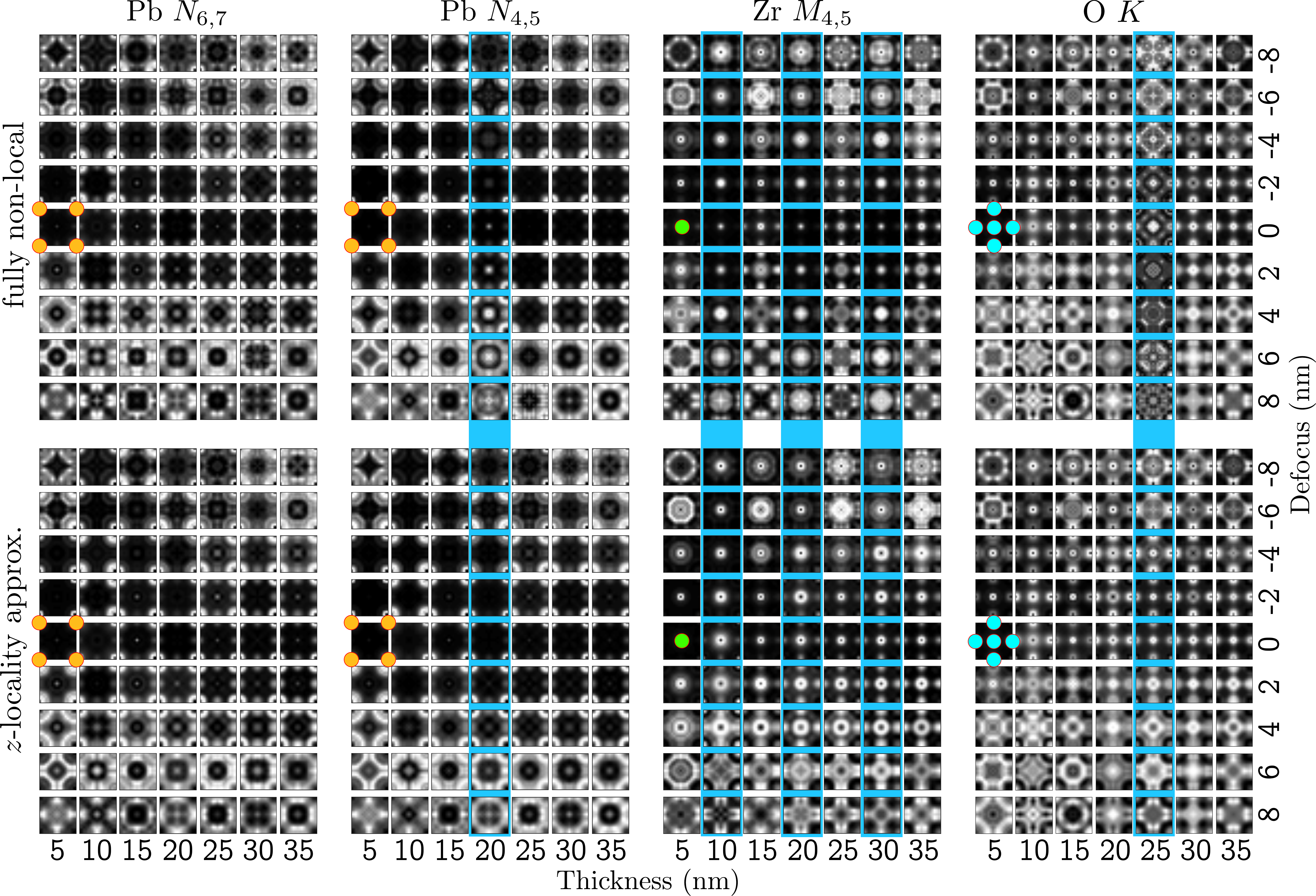}
  \caption{Defocus series of EF-HRTEM images of PbZrO$_3$, calculation for the Pb edges $N_{6,7}$ and $N_{4,5}$, Zr $M_{4,5}$ edge and O $K$ edge. The upper panels show calculations with full non-locality of the inelastic interaction, while the lower panels show calculations with $z$-locality approximation. The range of defoci is from $-8$~nm to $+8$~nm, as in Fig.~\ref{fig:sto110}. The thicknesses vary from 5~nm to 35~nm with 5~nm steps. Positions of atomic columns are marked by orange (Pb), green (Zr) and light blue (O) spheres in panels with zero defocus and thickness of 5~nm. Dimensions of each individual plot are 4.18~\AA{} $\times$ 4.18~\AA{}. Blue frames mark thicknesses, where the largest differences between fully non-local and $z$-local treatment can be observed, see text for more details.}
  \label{fig:pzo}
\end{figure*}

The oscillating character of the deviations between the full non-local calculations and the $z$-locality approximation, on the other hand, may be better explained in position space representation. Fig.~\ref{fig:channeling} shows the result of elastic scattering simulations, namely the well-known periodic channeling effect with the wave length of the oscillations depending on the weight of the columns. Note that the maximal channeling at the Ti column takes place at approximately 10 and 19 nm, which corresponds to thicknesses of large errors in the $z$-locality approximation. In other words, the neglection of 3D effects in the matrix element of the $z$-locality approximation appears to be most severe, when the intensity of the beam is concentrated at the inelastic scattering site. The effect might be related to the increased overlap of the beam electron wave function and the atomic wave function at strong channeling conditions, also increasing the impact of local variations of the wave functions within the overlap region. To foster our understanding we consider an additional example containing atomic species with a larger range of atomic numbers in the next section.

\subsection{Lead zirconate}

To probe, how the impact of the $z$-locality approximation depends on the weight of elements, we have performed simulations of PbZrO$_3$, which contains a rather heavy element Pb with atomic number 82, more than twice the atomic number of Sr in SrTiO$_3$. PbZrO$_3$ should therefore scatter the beam electrons much more strongly.

We simulated EF-HRTEM images of the $N_{6,7}, N_{4,5}$ edges of Pb, the $M_{4,5}$ edge of Zr, and the $K$ edge of O in a $[001]$ zone axis orientation. A parallel electron beam of kinetic energy 200~keV impinges along the $c$-axis on the crystal of PbZrO$_3$, which has a cubic structure with lattice parameter of 4.18~\AA{}. The unit cell was sampled on a real space grid of $64 \times 64 \times 60$. The reciprocal wave was the same as in the SrTiO$_3$ calculations above -- a convergent probe with convergence semi-angle of 37.5~mrad. The supercell for computing the reciprocal probe was $12 \times 12$ and considered thicknesses range from 5 to 35~nm sampled with steps of 5~nm, which corresponds to 12 unit cells. All aberrations of the imaging optics were set to zero, except for the defocus, which was varied in the same range as for SrTiO$_3$ above, i.e., from -8 to +8~nm with steps of 2~nm.

The results are summarized in Fig.~\ref{fig:pzo}. EF-HRTEM images for Pb edges show a rather weak impact of the $z$-locality approximation. The only exception is a sizable difference in the predicted intensities at the Zr column at 20~nm in the $N_{4,5}$ calculations. Note that this is also the sole appearance of a strong ZrO column excitation (visible in focus) in that edge. Although we again observe some sharper features and an increased dechanneling in nonlocal calculations, overall the impact of the $z$-locality approximation appears to be reduced, when compared to the Ti-$L_3$ edge calculations in SrTiO$_3$ above.

Interestingly, the calculations of the Zr-$M_{4,5}$ edge show again sizable differences. Note for example the volcano-shaped images around Zr atomic columns in the $z$-locality approximation, which often get ``filled'' in a fully-delocalized calculation, particularly at thicknesses of 10, 20 and 30~nm. Moreover, the intensity at the adjacent O column is larger than in the above Pb excitations. Overall, the situation for the Zr-$M_{4,5}$ edge resembles the SrTiO$_3$ calculations above. However, the channeling maxima at the Zr column appear at approximately 4~nm, 11~nm, 18~nm, 24~nm, and thus the correlation with the strength of $z$-nonlocality effects is less clear.

Finally, the situation with the oxygen $K$-edge reminds of the case of Pb edges. The overall differences between $z$-local and fully nonlocal calculations is rather small, except for one specific thickness of 25~nm. According to Fig.~\ref{fig:channeling}, oxygen columns attract less of electrons compared to atomic columns containing Zr or Pb elements. The period of oscillations, if any, must be larger than 30~nm. Yet, there is a maximum intensity on the oxygen column just around 25~nm. This fits with the previous argumentation, though the maximum here is very broad and questions arise, why the effect of $z$-locality is so pronounced at 25~nm, while it is visually almost non-existent at nearby thicknesses of 20~nm and 30~nm.

In summary, we do not observe any obvious correlations of the impact of $z$-locality approximation with the energy-loss of the edge or the mass of excited element. In several cases, also including SrTiO$_3$, the impact of $z$-locality approximation seems to correlate with the maxima of periodic channeling effects. This might hint towards a qualitative explanation of the differences. However, full explanation will most likely require analysing simultaneously both the incoming and the reciprocal wave and their relative properties.

\section{Conclusions and Outlook}

We have compared the Yoshioka equations based forward propagating approaches to reciprocal wave approaches for calculating inelastic scattering cross-sections. Their equivalency was demonstrated and their advantages and disadvantages have been discussed. In the reciprocal wave approach, we have shown how we can efficiently evaluate fully non-local inelastic scattering. Comparison of fully non-local calculations to calculations with $z$-locality approximation have uncovered the limitations of the latter approach. We often observe sizable and periodically changing differences between the two approaches, if the considered transitions stem from atomic columns subject to strong channeling conditions. This suggests that the fully non-local computations should be done for high-resolution zone axis conditions (e.g., atomic resolution EFTEM or Spectrum Imaging), whereas the $z$-locality approximation is acceptable for medium spatial resolution EELS measurements in out-of-zone axis (non-channeling) conditions. Moreover, we observe an amplification of the deviation patterns if large spatial aberrations (defocus, spherical aberration) contribute to the imaging process in EF-HRTEM. To provide a full account of the effect, further systematic studies considering inelastic transitions in a larger class of different materials are required. These studies should also address EFDIF patterns as well as EELS and EDX. We believe that our initial results regarding the impact of $z$-locality will stimulate further research efforts, including detailed comparisons with experiments.

\begin{acknowledgments}
J.R. acknowledges the Swedish Research Countil and the G\"{o}ran Gustafsson's Foundation for financial support. The simulations were performed at the National Supercomputing Centre at Link\"{o}ping University, under the Swedish National Infrastructure for Computing (SNIC). J.S. acknowledges the Center of Interdisciplinary Mathematics at Uppsala University. A.L. acknowledges funding from the European Research Council (ERC) under the European Union’s Horizon 2020 research and innovation programme (grant agreement No 715620).
\end{acknowledgments}

\end{document}